\documentclass[a4paper,11pt]{article}
\usepackage{pos}

\title{Run 3 performance and advances in heavy-flavor jet tagging in CMS}
\ShortTitle{Heavy-flavor jet tagging in CMS}

\author*[a]{Uttiya Sarkar on behalf of the CMS collaboration}

\affiliation[a]{III. Physikalisches Institut A, RWTH Aachen,\\
  Sommerfeldstr. 16, 52074 Aachen, Germany}

\emailAdd{uttiya.sarkar@cern.ch}

\abstract{Identification of hadronic jets originating from heavy-flavor quarks is extremely important to several physics analyses in High Energy Physics, such as studies of the properties of the top quark and the Higgs boson, and searches for new physics. Recent algorithms used in the CMS experiment were developed using state-of-the-art machine-learning techniques to distinguish jets emerging from the decay of heavy flavour (charm and bottom) quarks from those arising from light-flavor (udsg) ones. Increasingly complex deep neural network architectures, such as graphs and transformers, have helped achieve unprecedented accuracies in jet tagging. New advances in tagging algorithms, along with new calibration methods using flavour-enriched selections of proton-proton collision events, allow us to estimate flavour tagging performances with the CMS detector during early Run 3 of the LHC.}

\FullConference{42nd International Conference on High Energy Physics (ICHEP2024)\\
18-24 July 2024\\
Prague, Czech Republic\\}

\begin{document}
\maketitle

\section{Introduction}
Heavy flavour tagging is of utmost importance in the CMS experiment~\cite{Chatrchyan:2008aa,CMS:2017wtu,cms2018} at the Large Hadron Collider (LHC) for identifying jets originating from heavy quarks, particularly bottom (b) and charm (c) quarks. This method plays a pivotal role in various physics analyses, including those focused on the Standard Model (SM), top quark physics, Higgs boson production, and Beyond the Standard Model (BSM) searches, such as Supersymmetry (SUSY) processes. 

To differentiate heavy-flavor jets from light-flavor quark or gluon jets, standard methods make use of detector inputs from reconstructed charged particle tracks and information about secondary vertices reconstructed within jets. The jet-related data is incorporated into advanced machine-learning techniques. Although there is a rapid improvement in tagging methods, it is essential to ensure the consistency between data and simulations since the detector conditions change over time and simulations are often imperfect. 

This proceeding briefly discusses about the various tagging strategies that are used in CMS focusing on the most recently developed tagging algorithm, namely UnifiedParticleTransformer (UParT)~\cite{CMS-DP-2024-066}. It then presents a comparison of data and simulation and derived scale factors using the
proton-proton collision data collected by the CMS detector at $\sqrt{s}$ =13.6 TeV during 2022-2023 data-taking periods~\cite{Hayrapetyan_2024}. Then it shows the performance and validation results for large-radius jets. Finally, it concludes by discussing the two modern computing frameworks used for training and commissioning purposes.

\section{Heavy-flavor tagging in CMS}
Heavy-flavor jet tagging depends on variables related to the characteristics of heavy-flavor hadrons within jets, such as the presence of secondary vertices, higher track multiplicities, and more tracks with positively signed impact parameters.

\begin{figure}
    \centering
    \includegraphics[width=0.44\linewidth]{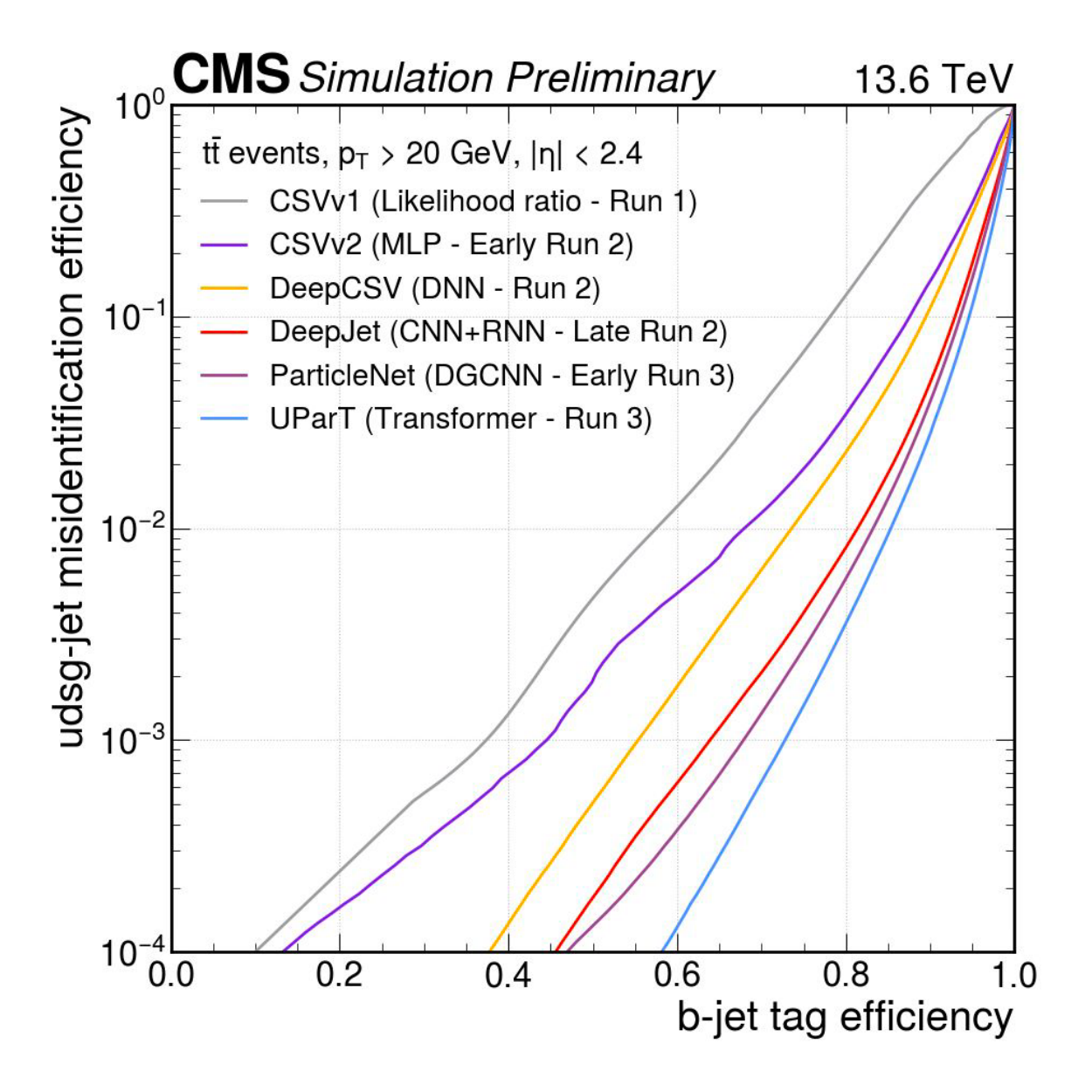}
    \includegraphics[width=0.55\linewidth]{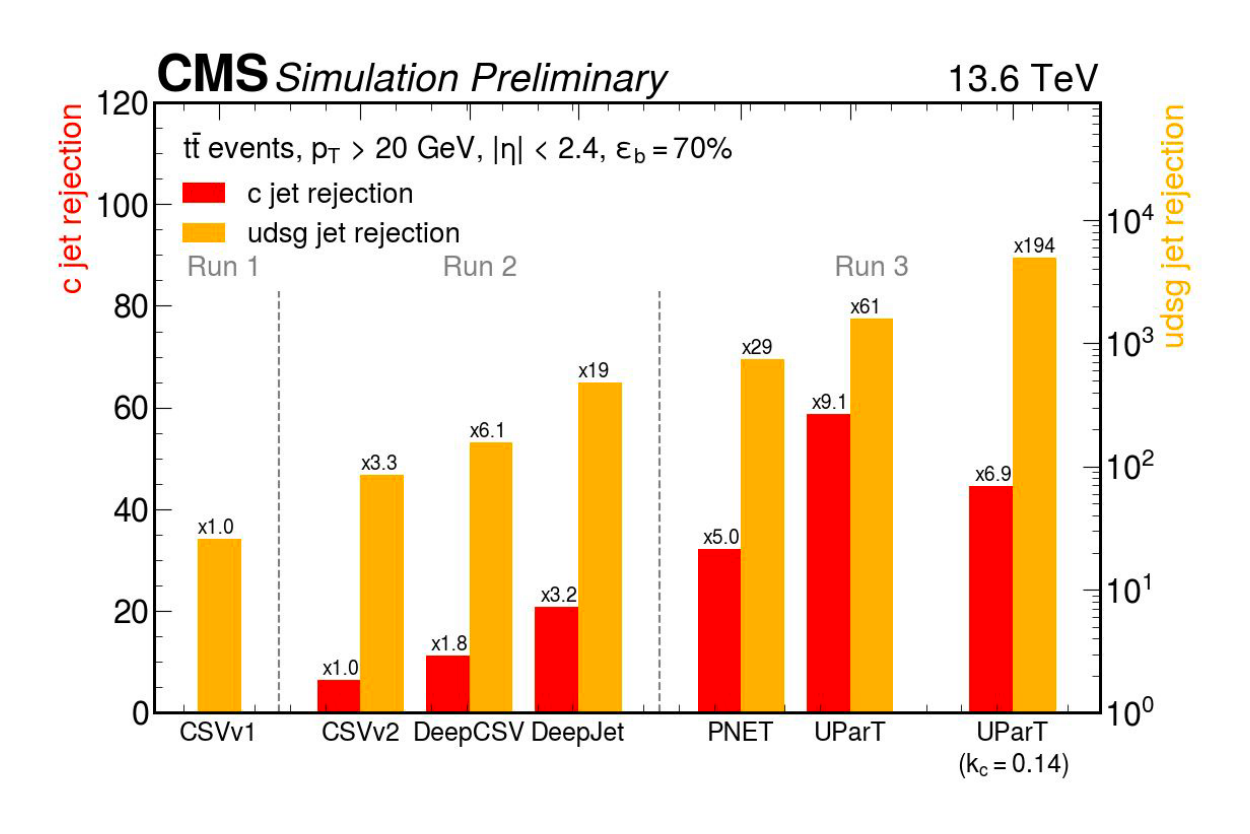}
    \caption{Evolution of the b-tagging performance for jet flavor identification algorithms used in CMS from Run 1 to Run 3. Left plot shows the increase in tagging performance follows the progression in machine learning whereas the right plot shows that the light- (udsg, yellow bars) and c-jet (red bars) rejection for a fixed b-jet identification efficiency of $70\%$ for taggers from Run 1 to Run 3. }
    \label{fig:01}
\end{figure}

The historical evolution of flavor taggers in the CMS experiment reflects significant advancements in machine learning techniques and algorithms over the past decade. Starting with the combined secondary vertex (CSV)~\cite{Weiser:2006md} method, which relied on a likelihood ratio during Run 1, CMS transitioned to more sophisticated models, such as CSVv2 (Multi-Layer-Percrepton) during early Run 2. By 2017, CMS introduced DeepCSV~\cite{PhysRevD.94.112002}, utilizing deep neural networks (DNN), followed by DeepJet~\cite{Bols_2020} in late Run 2, which combined convolutional neural networks (CNN) and recurrent neural networks (RNN) for enhanced performance. The introduction of ParticleNet~\cite{Qu:2019gqs} in early Run 3 further improved tagging capabilities by treating particles as unordered clouds, using the dynamic graph convolutional neural network (DGCNN) architecture. The latest advancements in Run 3, such as RobustParticleTransformer (RobustParT)~\cite{Stein:2022nvf,CMS-DP-2022-050} and UParT~\cite{CMS-DP-2024-066}, leverage transformer-based architectures --- a multi-head attention mechanism with pairwise interaction features between all input jet constituents and secondary vertices~\cite{Qu:2022mxj}. These models also incorporate input feature distortions (adversarial attacks) to enhance robustness. Overall, the evolution of these algorithms has led to substantial improvements in tagging efficiency, as demonstrated in figure~\ref{fig:01} showing reduced light-flavour mistagging rates at different b-jet tagging efficiencies (left) and improved light or c-jet rejection at a fixed b-jet tagging efficiency of 70\% (right). The last generation models not only tag b and c jets but are also capable of tagging strange (s) and hadronic $\tau$ ($\tau_h$) jets.

To evaluate the performance of b-jets and c-jets, several discriminators—such as \texttt{BvsAll}, \texttt{BvsL}, \texttt{CvL}, and \texttt{CvB}—are defined based on the probability of identifying different jet types. These transformations allow discriminating specifically between bottom (\texttt{b}), charm (\texttt{c}), strange (\texttt{s}), light (\texttt{ud}), and gluon jets. The definitions are:\\
\centerline{\texttt{BvsAll} = {\texttt{prob(b)}} / {[1-\texttt{prob(b)}]},} \\
\centerline{\texttt{BvsL} = {\texttt{prob(b)}} / {[\texttt{prob(b)} + \texttt{prob(udsg)}]},} \\
\centerline{\texttt{CvsL} = {\texttt{prob(c)}} / {[\texttt{prob(c)} + \texttt{prob(udsg)}]}, and} \\
\centerline{\texttt{CvsB} = {\texttt{prob(c)}} / {[\texttt{prob(c)} + \texttt{prob(b)}]}}
\\For \texttt{s} and \texttt{$\tau_h$} tagging, a broader classification is used, defined as:\\
\centerline{\texttt{svsX} = {\texttt{[prob(s)]}} / {\texttt{[prob(s)} + \texttt{prob(X)}]}}\\
\centerline{\texttt{$\tau_h$vsX} = {\texttt{[prob($\tau_h$)]}} / {\texttt{[prob($\tau_h$)} + \texttt{prob(X)}]}}
where X is any other jet type.

\section{UParT}
UParT is based on the ParticleTransformer architecture~\cite{Qu:2022mxj} specifically designed for AK4 jets (anti-$k_{T}$ jets with a cone radius of 0.4). It integrates heavy-flavor identification with flavor-aware jet energy regression and jet energy resolution estimation in a comprehensive manner. The model introduces pairwise interaction features between jet constituents and secondary vertices, enhancing its capability to capture complex internal jet relationships. A key advancement in UParT is the use of a novel Adversarial Training approach. This enables the model to learn from input feature distortions, significantly improving its robustness against mismodeling in simulations. Furthermore, UParT preserves feature importance mapping through adversarial attack gradients, highlighting critical features for jet classification, ensuring both high performance and robustness in heavy-flavor tagging. 

\begin{figure}
    \centering
    \includegraphics[width=0.44\linewidth]{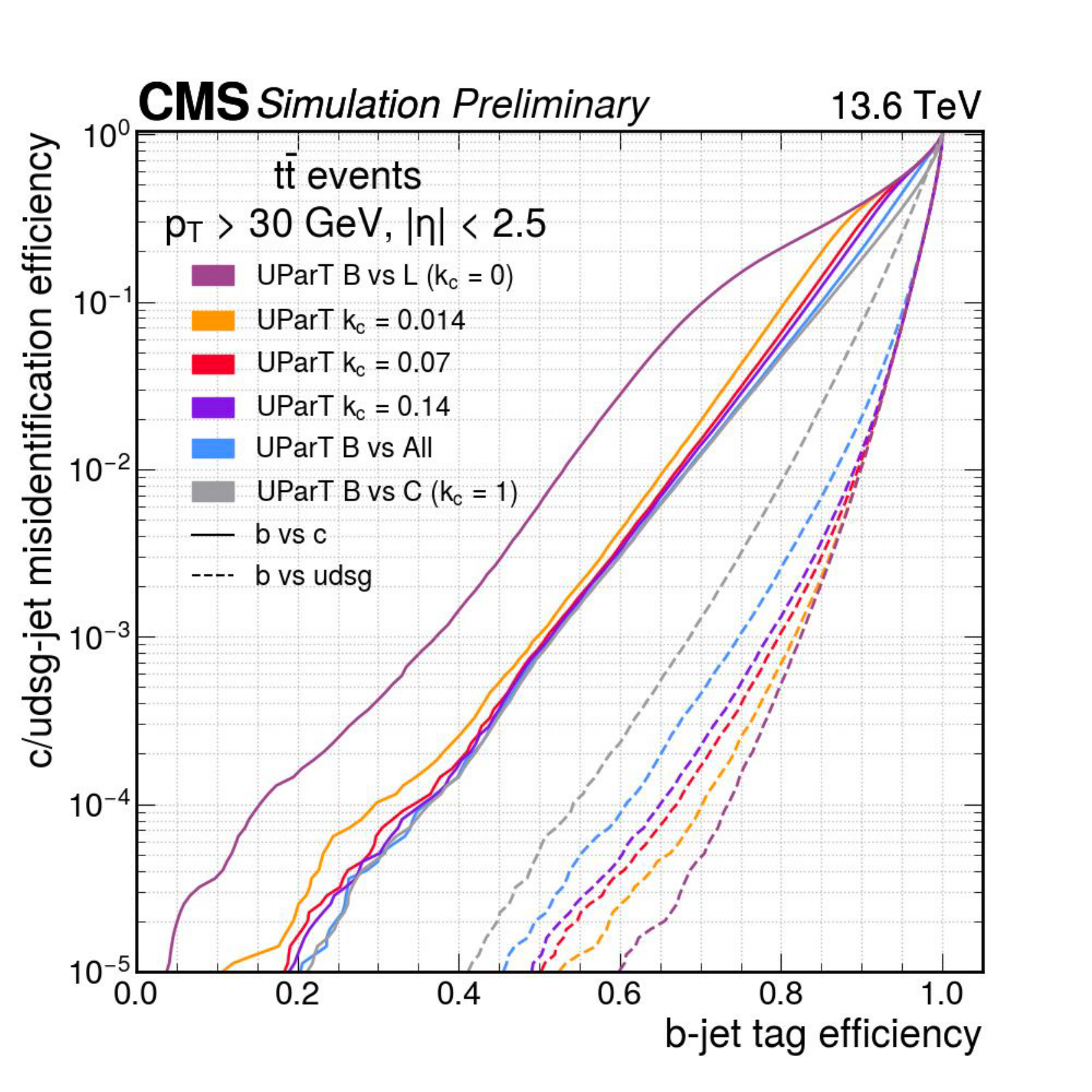}
    \includegraphics[width=0.44\linewidth]{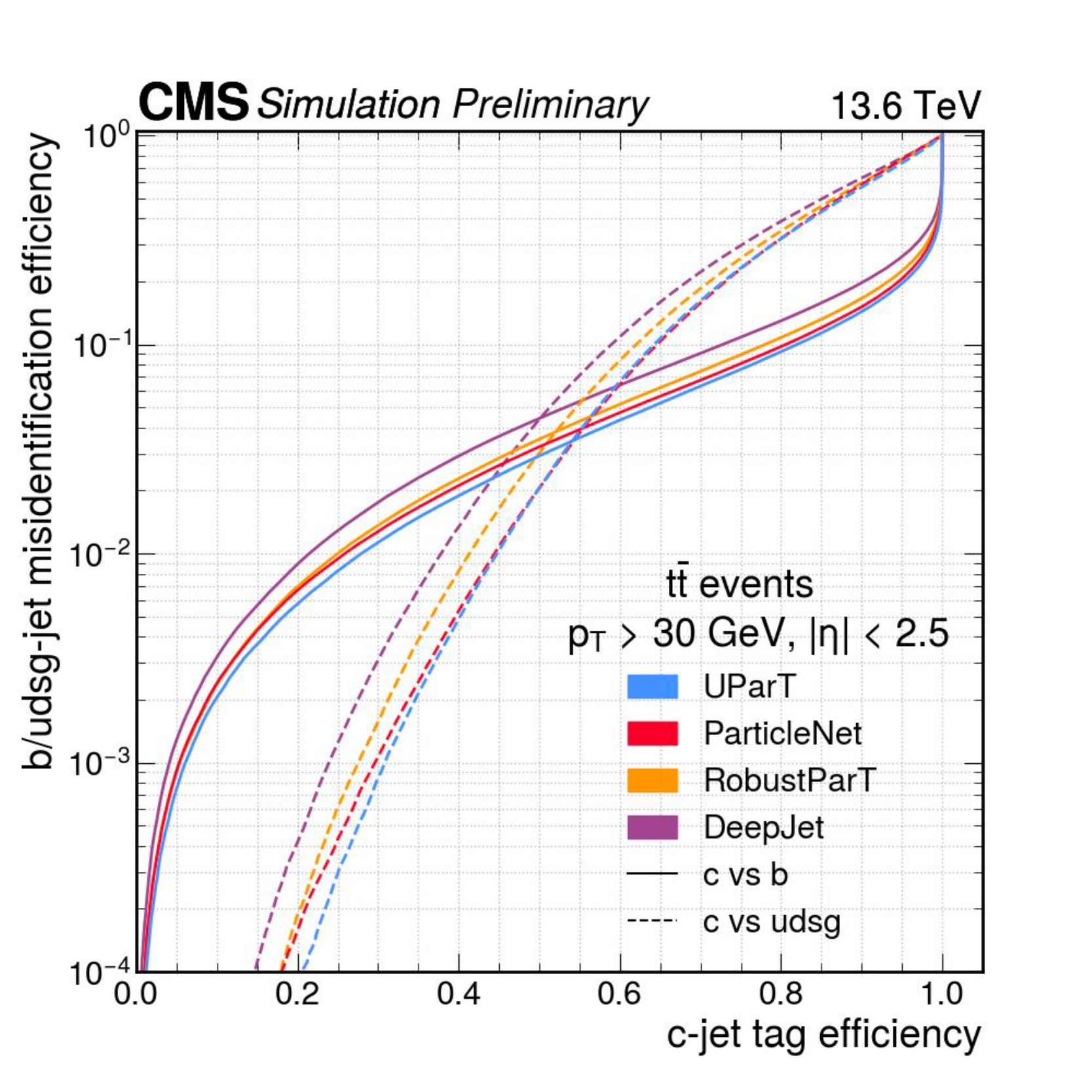}
    \caption{Left: ROC curves of post-training reweighted discriminators for b-tagging. Both BvsL and BvsC shows the best rejection against their specific flavor but fail to achieve a good rejection for the other. The different weighted BvsAll show a trade-off for BvsC and BvsL showing a significant improvement in BvsL for a limited impact on BvsC performance. Right: c-tagging ROC curves. UParT shows state-of-the-art performance for both b and light jet rejections.}
    \label{fig:02}
\end{figure}

The left plot in figure~\ref{fig:02} shows the c/udsg jet misidentification rate vs. the b-jet tagging efficiency for jets with $\mathrm{p_T} > 30$ GeV and $|{\eta}| < 2.5$ taken from a simulated ttbar sample. Whereas, the right plot in figure~\ref{fig:02} shows the b/udsg jet misidentification efficiency versus the c-jet tagging efficiency for different tagging algorithms in the same jet selection. A consistent improvement in the tagging performance is observed with UParT being the most performing model so far in c-jet identification.  

\begin{figure}
    \centering
    \includegraphics[width=0.44\linewidth]{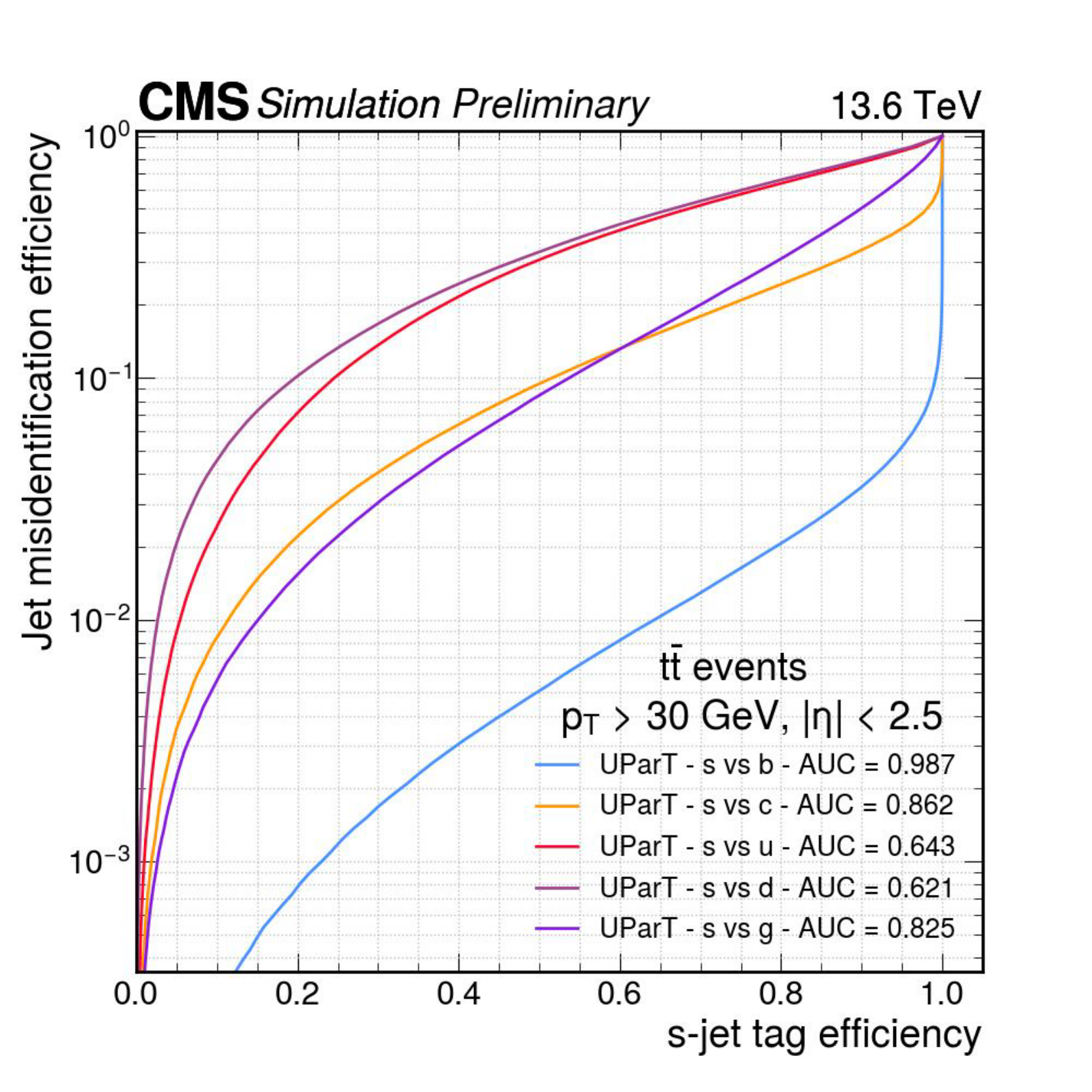}
    \includegraphics[width=0.44\linewidth]{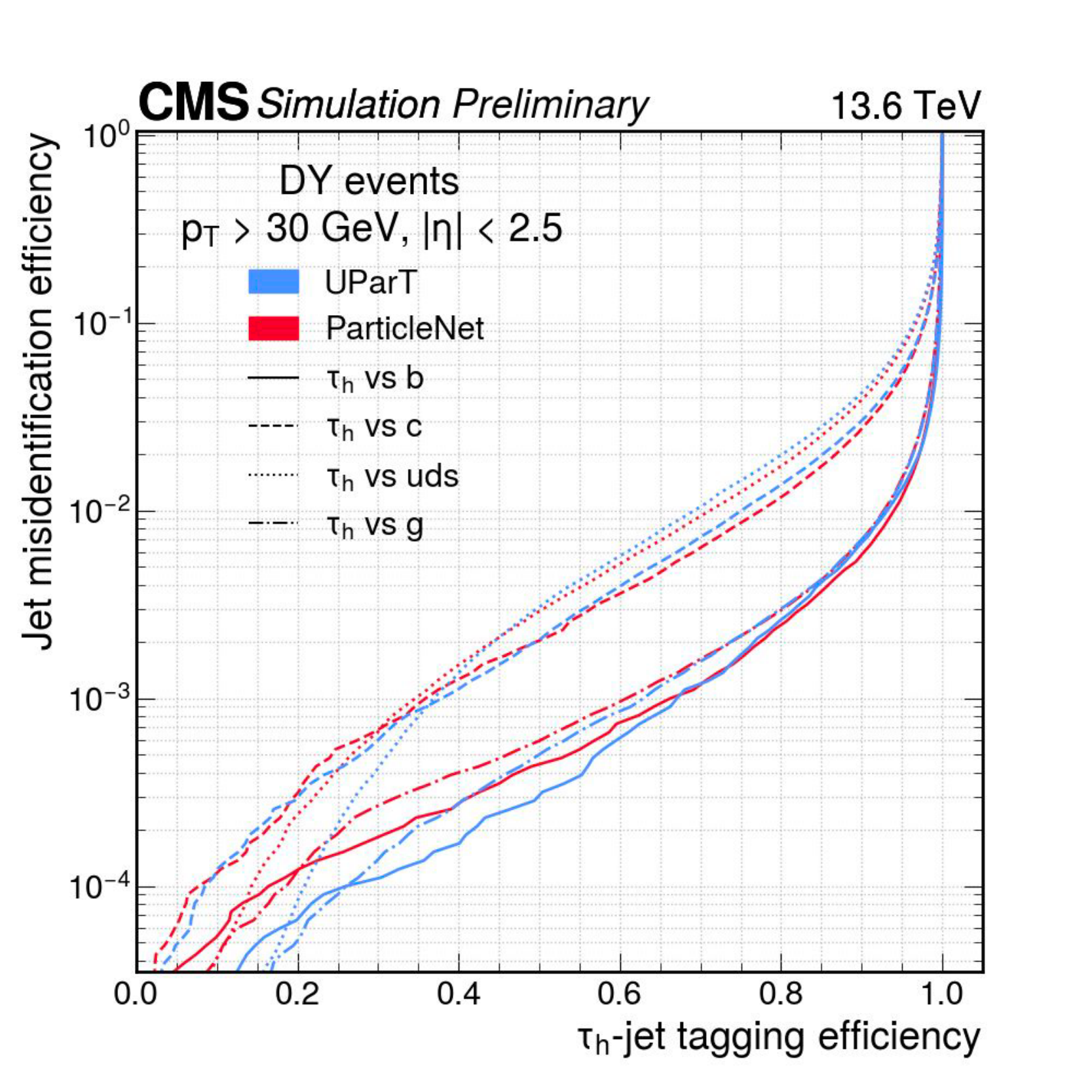}
    \caption{Left: First s-tagging ROC curves in CMS experiment. Performances indicates we can achieve a low efficiency s-tagger. Right: $\tau$-tagging ROC curves. ParticleNet and UParT show similar performances. ParticleNet performs better at high misidentification rate and UParT at lower rate.}
    \label{fig:03}
\end{figure}

Additionally, UParT extends its capabilities by introducing an s-jet classifier, making it the first attempt to identify jets originating from s-quarks within the CMS experiment. The left plot in figure~\ref{fig:03} shows the s-jet tagging efficiency against the misidentification rate of other jets. It can already achieve ~20\% tagging efficiency for a mididentification probability of $10^{-3}$. In addition, UParT extends its capabilities to include hadronic tau classification, as demonstrated in the right plot of figure~\ref{fig:03}, further enhancing its versatility across a wide range of particle tagging tasks.

\section{Data vs. Simulations}
Discrepancies between data and simulations arise from multiple factors that impact the accuracy of jet tagging and classification in high-energy physics experiments. These mismatches are often attributed to imperfect modeling of input variables used in tagging algorithms, which translates into biases in the predicted distributions of jet-related observables, such as tagger scores or discriminator values. Calibration errors, especially in the detector’s alignment and tracking systems, further exacerbate these differences.

The misalignment between real-world data and simulated outputs manifests as shifts in peak positions and trends in tagger scores. Such mismatches are prominent before the implementation of scale factors (SFs), which are essential for correcting these discrepancies by adjusting the simulation results to better match the data. These corrections are necessary to maintain the validity of tagging algorithms in both inclusive and flavor-aware tasks.

\begin{figure}
    \centering
    \includegraphics[width=0.5\linewidth]{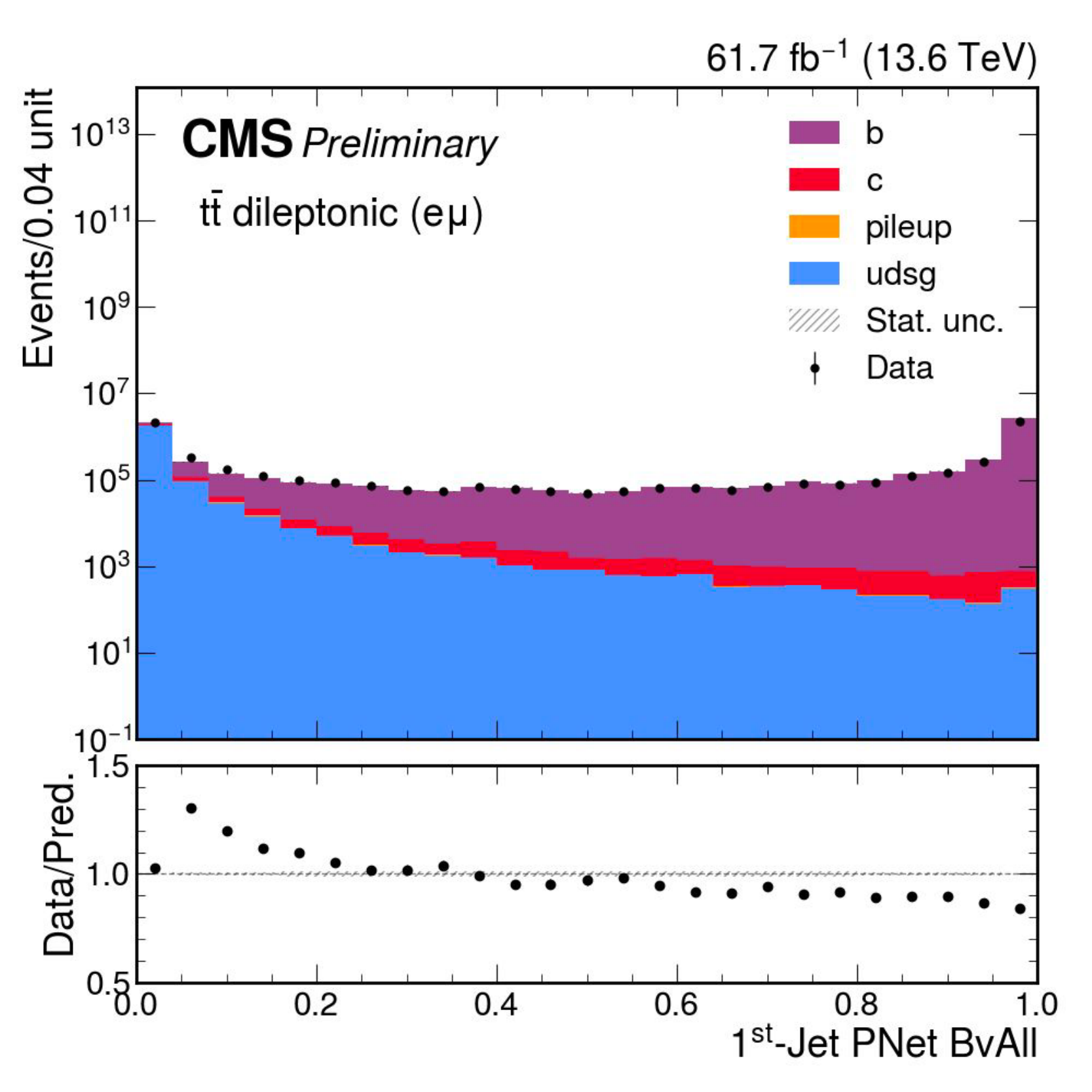}
    \caption{ParticleNetAK4 BvAll discriminator of the selected jet with highest $\mathrm{p_T}$. A lower tagger score in the mismatched peak position and downward trend are observed.}
    \label{fig:04}
\end{figure}

Figure~\ref{fig:04} presents the BvsAll discriminator distribution for the ParticleNet tagger in events enriched with $t\bar{t}$ decaying to dileptons, using data collected during the 2022-2023 run period, corresponding to an integrated luminosity of $61.7 , \mathrm{fb^{-1}}$~\cite{CMS-DP-2024-024}. The bottom panel shows the Data/Simulation ratio, which noticeably deviates from unity in the low score region.
\section{Calibration and Scale Factors}
As discussed in the previous section, the calibration of heavy-flavor jets in the CMS experiment is essential to improve the precision of tagging algorithms. This calibration is crucial for improving the accuracy of physics analyses that involve heavy-flavor jets. Scale Factors (SFs)~\cite{CMS-DP-2024-025} are introduced to correct discrepancies between the tagging efficiencies observed in data and those predicted by simulations. These SFs are derived by comparing the performance of tagging algorithms on data and simulations for various jet flavors (b, c, and light jets).

One commonly used method for deriving SFs involves calibrating b-tagging or c-tagging discriminators by analyzing datasets enriched with heavy-flavor jets, such as those produced in top quark pair decays (b enriched) or jets produced in association with W bosons (c enriched). During this process, the distributions of discriminants (e.g. BvsAll, ... CvsL and CvsB) are examined in both data and simulation. To achieve reliable SFs, systematic uncertainties, such as those due to differences in detector conditions or modeling inaccuracies, are accounted for, which ensures robust estimates.

SFs can be derived through two primary methods: working-point-based and shape-based calibration. In the working-point-based approach, SFs are calculated for specific thresholds (working points) of a tagging algorithm, providing a single correction factor for each working point. Shape-based calibration, on the other hand, adjusts the full distribution of a discriminant (e.g., the output score of a b-tagging algorithm) to improve agreement across a range of values. 

\begin{figure}
    \centering
    \includegraphics[width=0.5\linewidth]{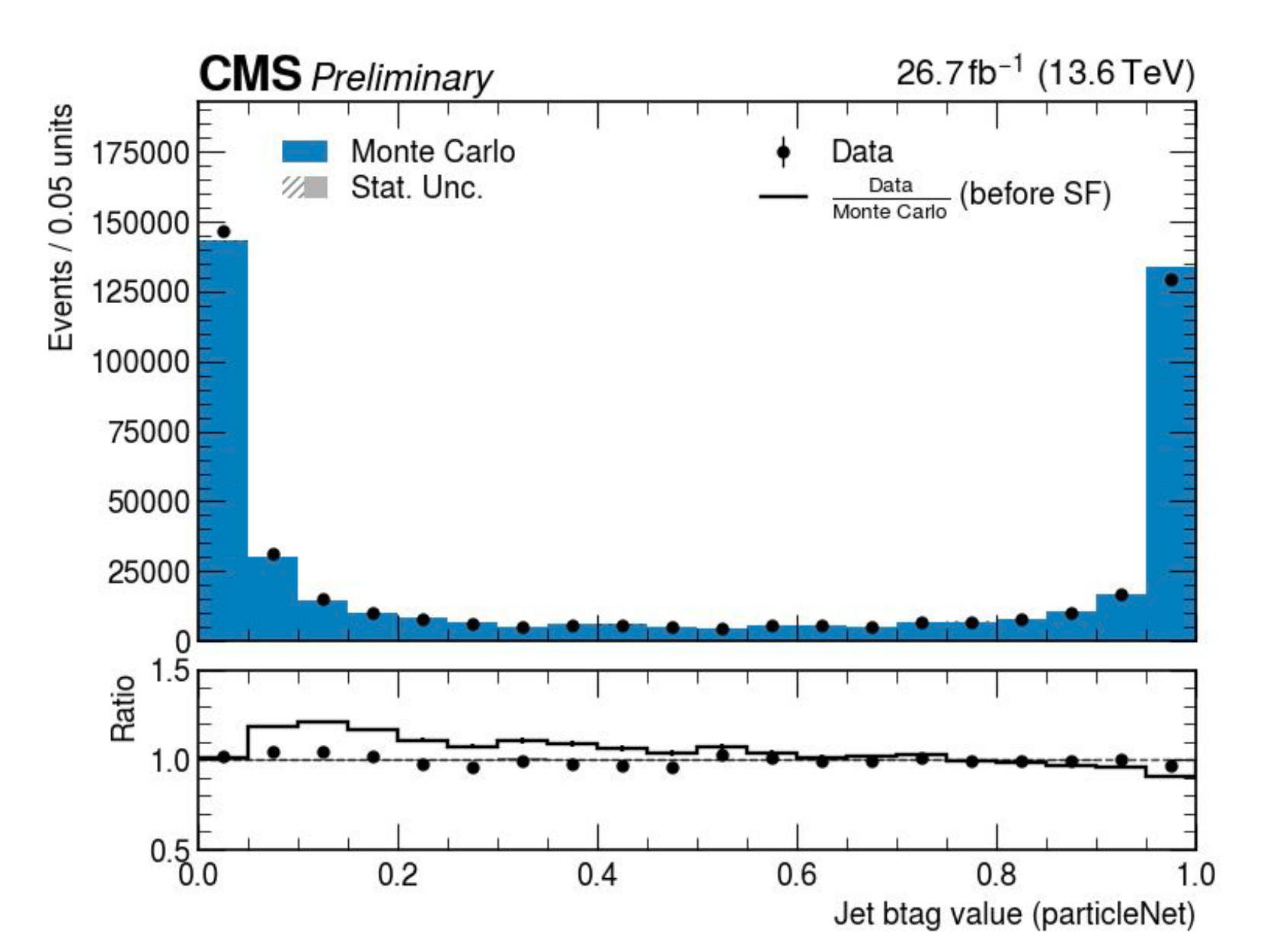}
    \includegraphics[width=0.44\linewidth]{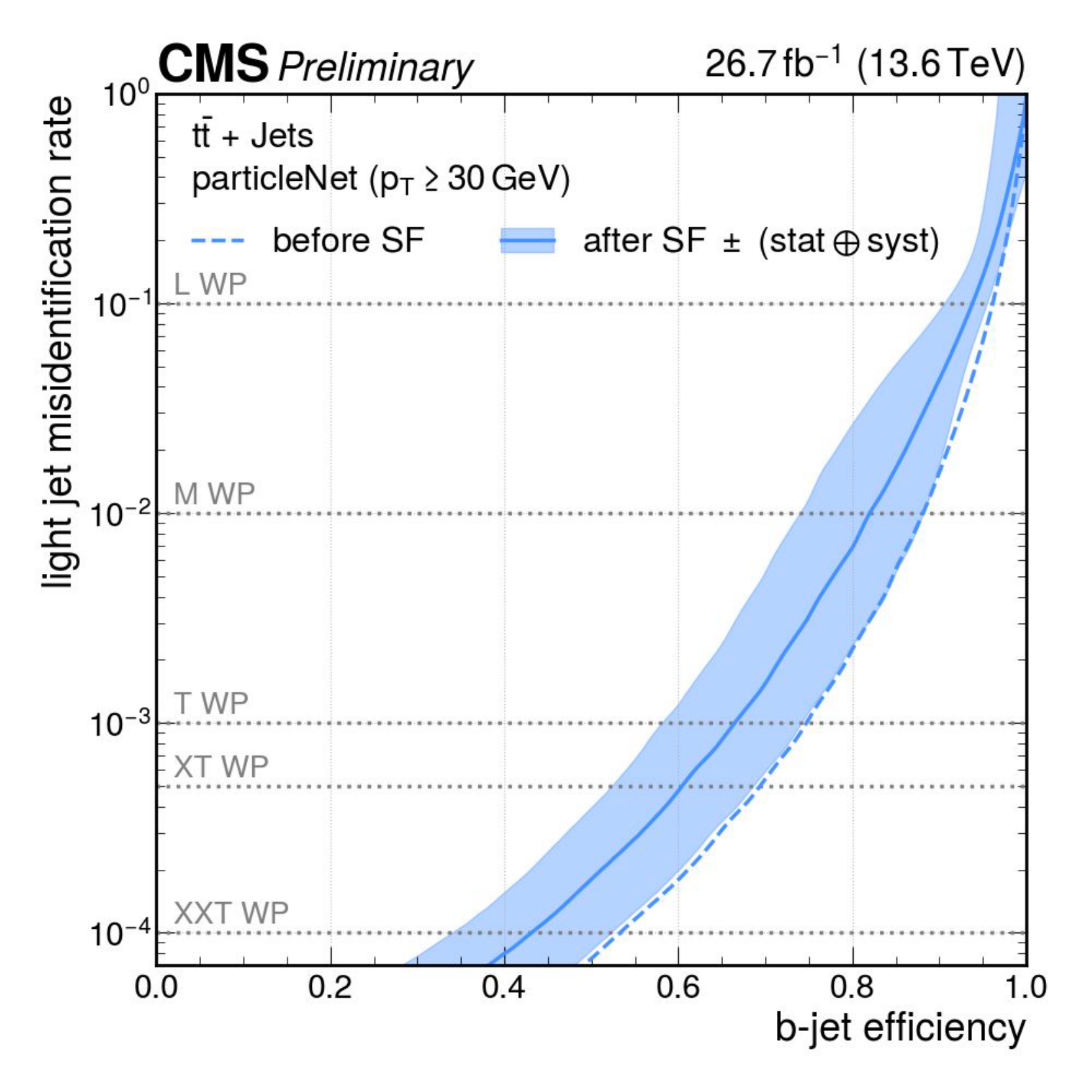}
    \caption{Left plot shows the distribution of the b jet discriminant score of the ParticleNet algorithm for jets the $e\mu$ + jets phase space for data recorded in late 2022 period. The recorded data and simulated events after applying the SFs are shown. In the lower panel the data-to-simulation ratio after the SF application is shown and in addition the ratio of data and simulation before the SF application is indicated. After the application of SFs, a better agreement is observed. Right plot shows the b jet identification efficiency versus light jet misidentification rate for the particleNet algorithm in late 2022 data evaluated for jets in hadronic $t\bar{t}$ events. The distributions are shown before and after scale factor application (including uncertainties) to match the efficiencies in data.}
    \label{fig:05}
\end{figure}

The comparison between pre- and post-calibration distributions is shown in figure~\ref{fig:05}. After applying the SFs, the agreement between data and simulation improves considerably, mitigating the effects of mismodeling. In the left plot, the b-tag discriminant values from ParticleNet for both data and simulations are compared before and after SF application, using data from late 2022 with an integrated luminosity of 26.7 $\mathrm{fb}^{-1}$. The right plot illustrates the the light-jet misidentification rate versus b-jet efficiency, where some deterioration in performance is observed post-calibration.

Thus, proper calibration of SFs is vital for reducing systematic uncertainties and ensuring reliable and consistent jet flavor tagging across different datasets.

\section{Merged jet tagging}
The tagging of boosted jets plays a crucial role in analyses focusing on highly-energetic decays of processes such as $H\rightarrow b\bar{b}$ and $H\rightarrow c\bar{c}$. In such cases, the Higgs boson decays into two jets that are often collimated, forming a single "boosted jet" due to its high momentum. To distinguish signal jets from background, several innovative tagging techniques have been developed to exploit the substructure of these jets.

Various machine learning (ML)-based algorithms have been employed to tag these jets. Some of the key algorithms used are:
\begin{itemize}
    \item \textbf{Double-b (DNN):} a deep neural network to differentiate jets containing bb-quarks from other jets,
    \item \textbf{DeepAK8MD (CNN):} A convolutional neural network (CNN) algorithm that is mass-decorrelated (MD) to mitigate mass dependence in tagging, 
    \item \textbf{DeepDoubleX (CNN, RNN):} combines CNN and recurrent neural networks (RNNs),
    \item \textbf{ParticleNetMD (DGCNN):} based on dynamic graph convolutional neural networks (DGCNN). 
\end{itemize}

\begin{figure}
    \centering
    \includegraphics[width=0.44\linewidth]{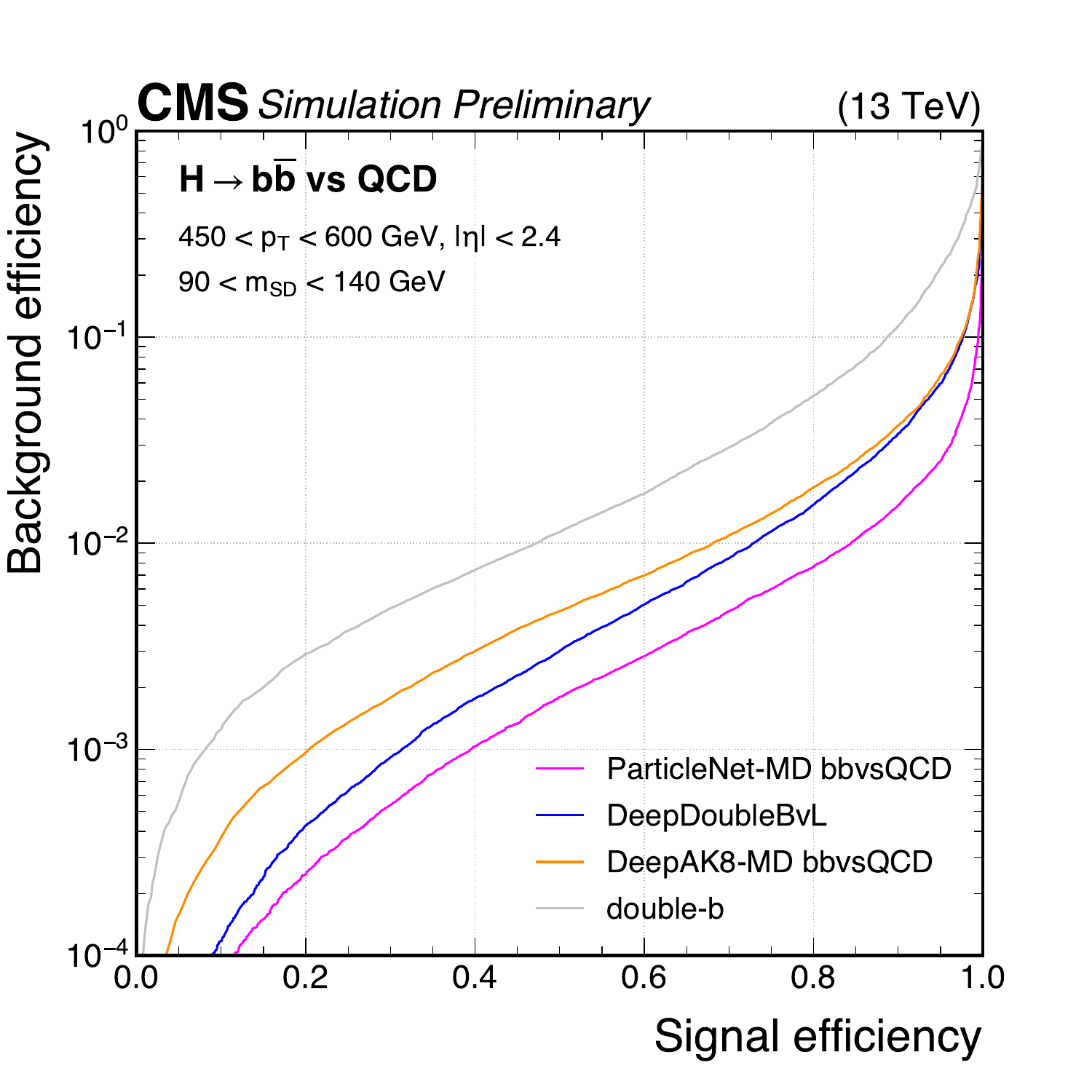}
    \includegraphics[width=0.44\linewidth]{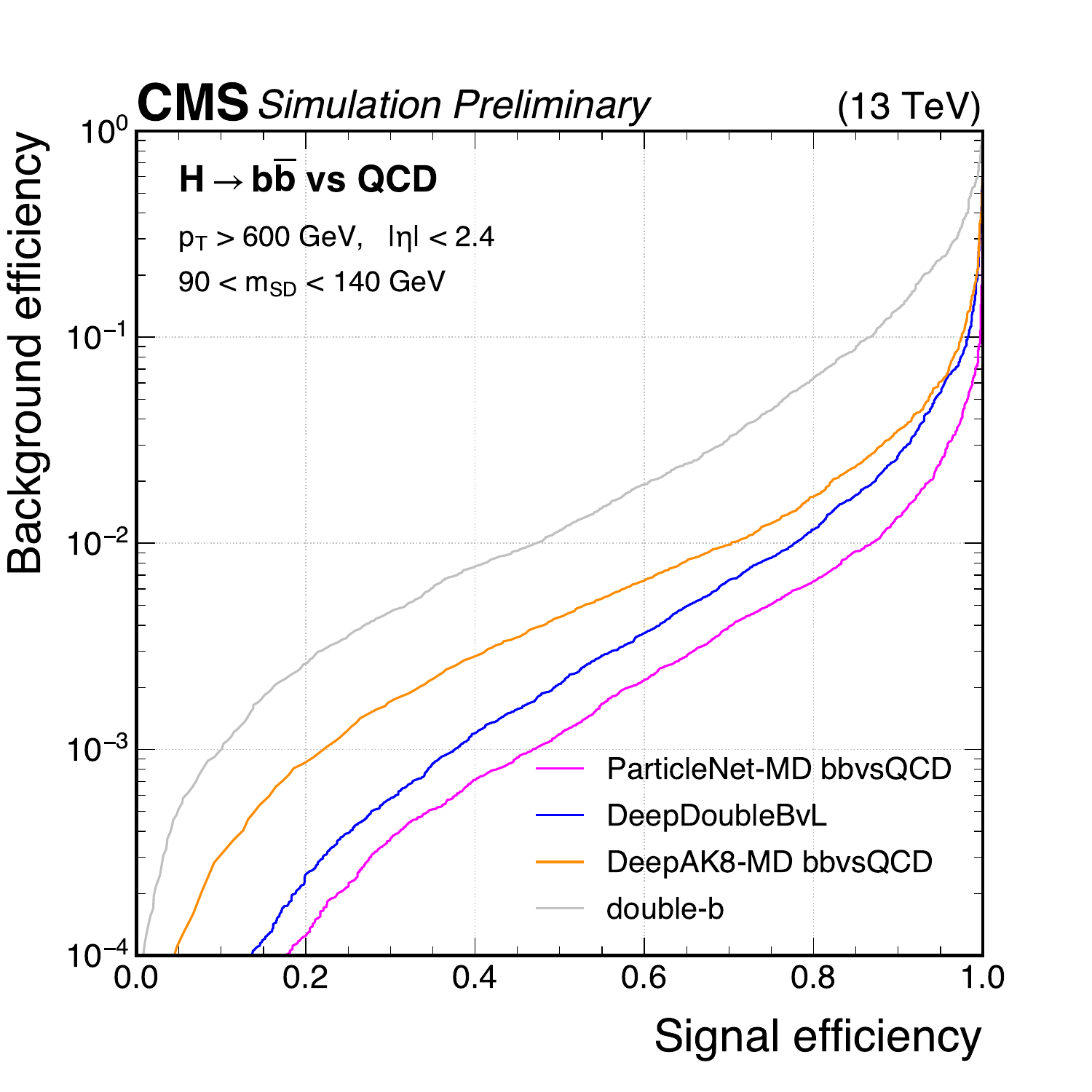}
    \caption{ROC curves for signal efficiency vs. background efficiency where signal is $H\rightarrow b\bar{b}$ and the background is inclusive QCD. The plot compares the performance of different boosted taggers in 450<$\mathrm{p_T}$<600 GeV (left) and $\mathrm{p_T}$ > 600 GeV (right) for a mass selection between 90 and 140 GeV. }
    \label{fig:06}
\end{figure}

\begin{figure}
    \centering
    \includegraphics[width=0.44\linewidth]{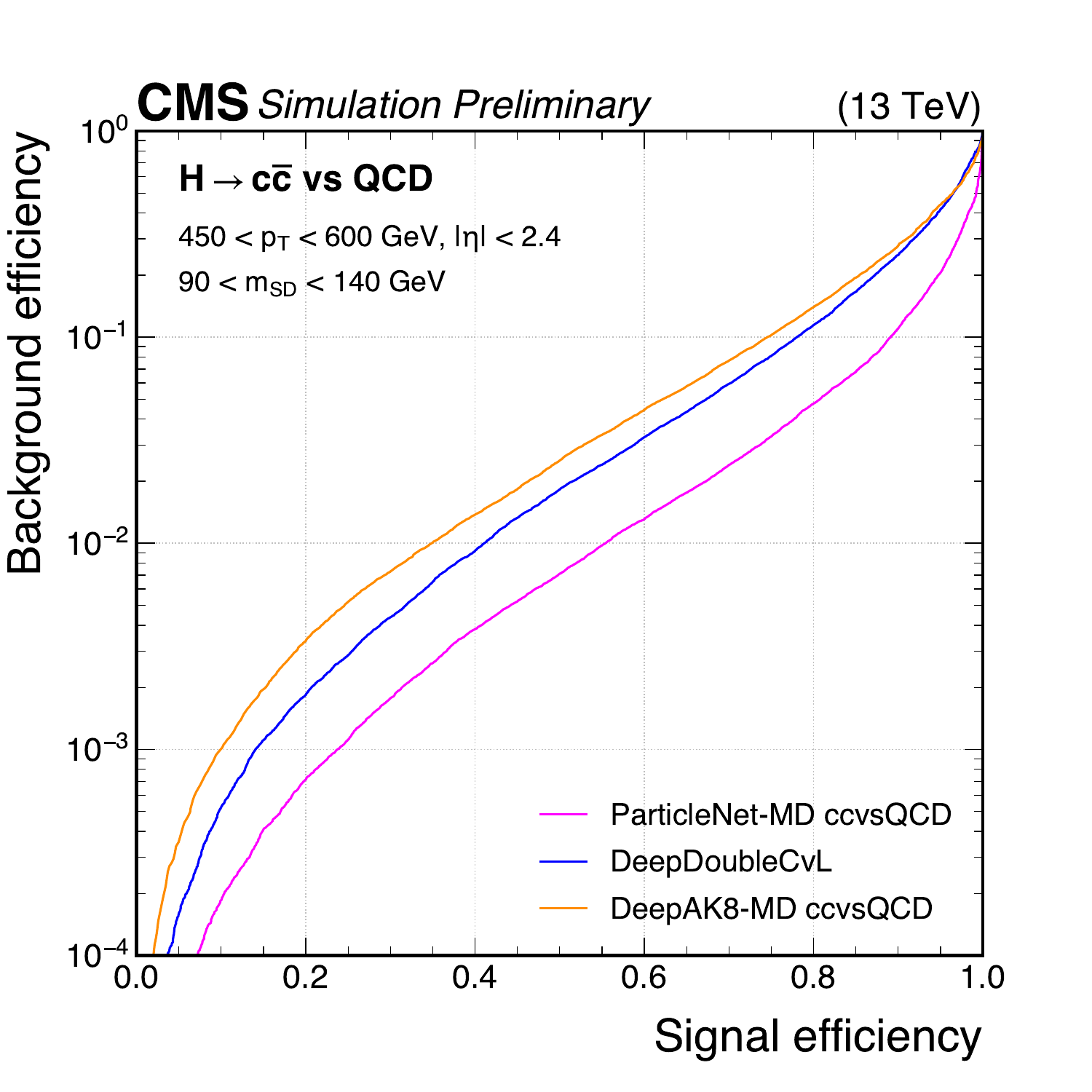}
    \includegraphics[width=0.44\linewidth]{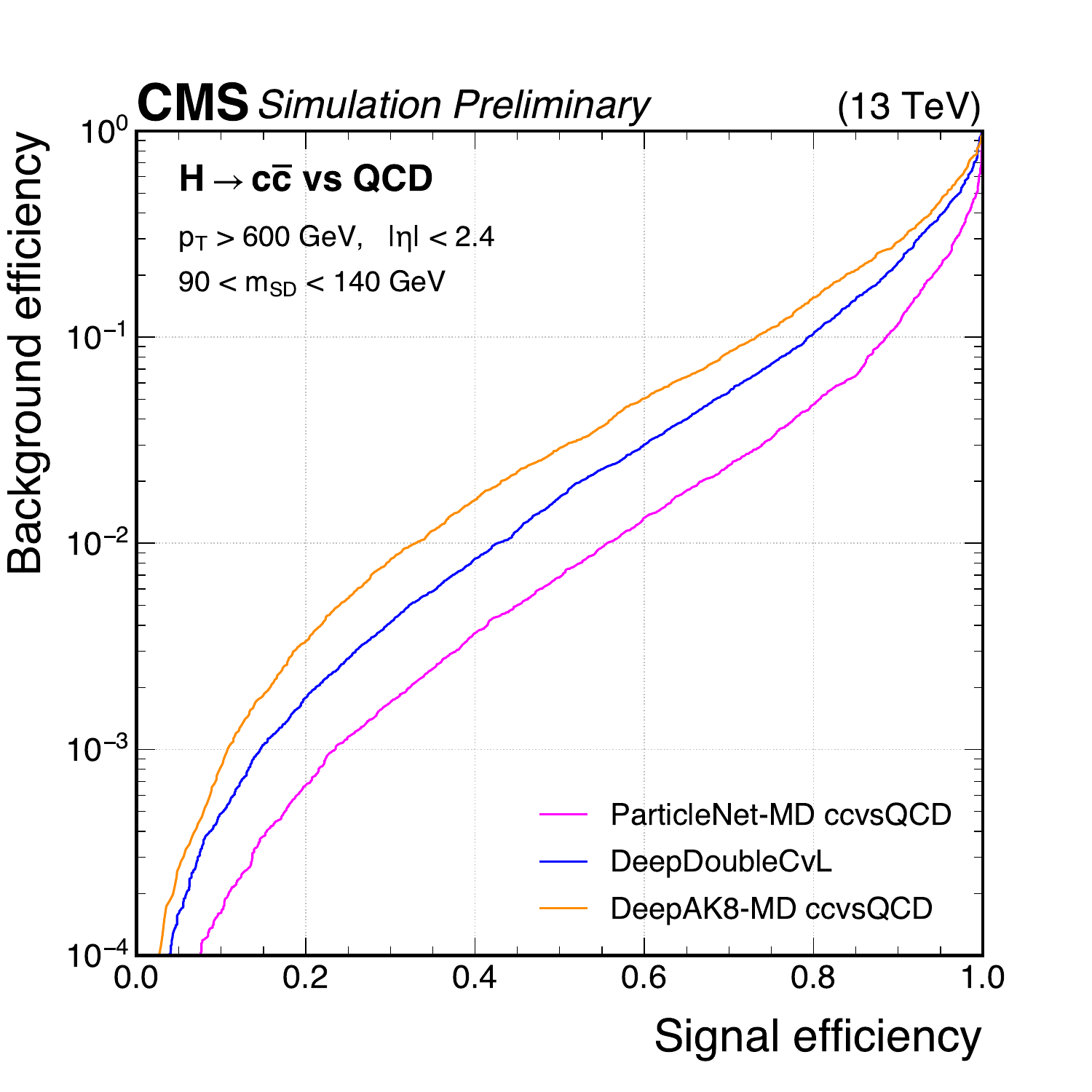}
    \caption{ROC curves for signal efficiency vs. background efficiency where signal is $H\rightarrow c\bar{c}$ and the background is inclusive QCD. The plot compares the performance of different boosted taggers in 450<$\mathrm{p_T}$<600 GeV (left) and $\mathrm{p_T}$ > 600 GeV (right) for a mass selection between 90 and 140 GeV.}
    \label{fig:07}
\end{figure}

The Run 2 performance of these algorithms is illustrated in figure~\ref{fig:06} for $H\rightarrow b\bar{b}$ and in figure~\ref{fig:07} for $H\rightarrow c\bar{c}$~\cite{CMS-PAS-BTV-22-001}. The signal efficiency versus background efficiency plot shows that the ParticleNetMD algorithm outperforms the older algorithms, offering superior discriminating power for both processes.

\begin{figure}
    \centering
    \includegraphics[width=0.4\linewidth]{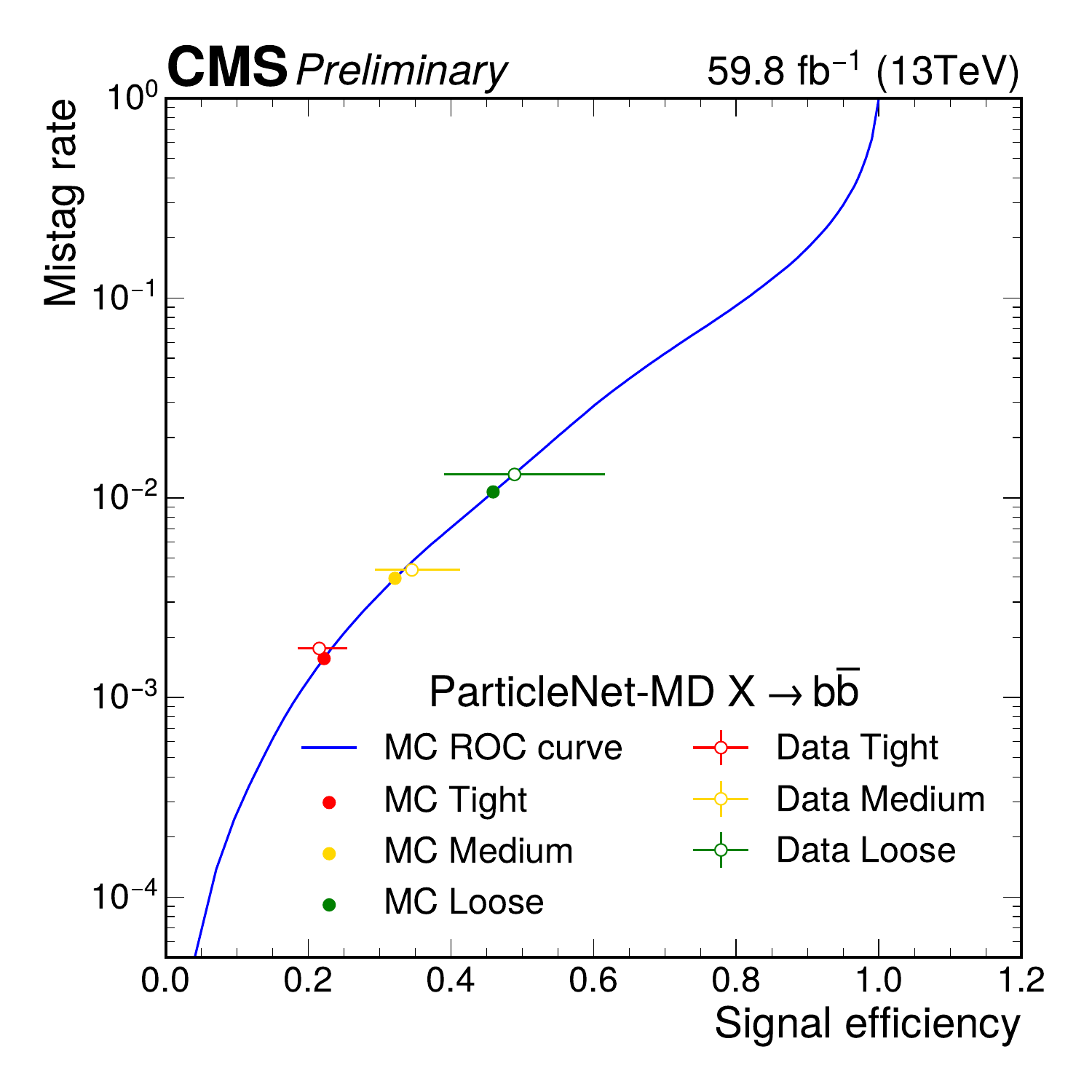}
    \includegraphics[width=0.52\linewidth]{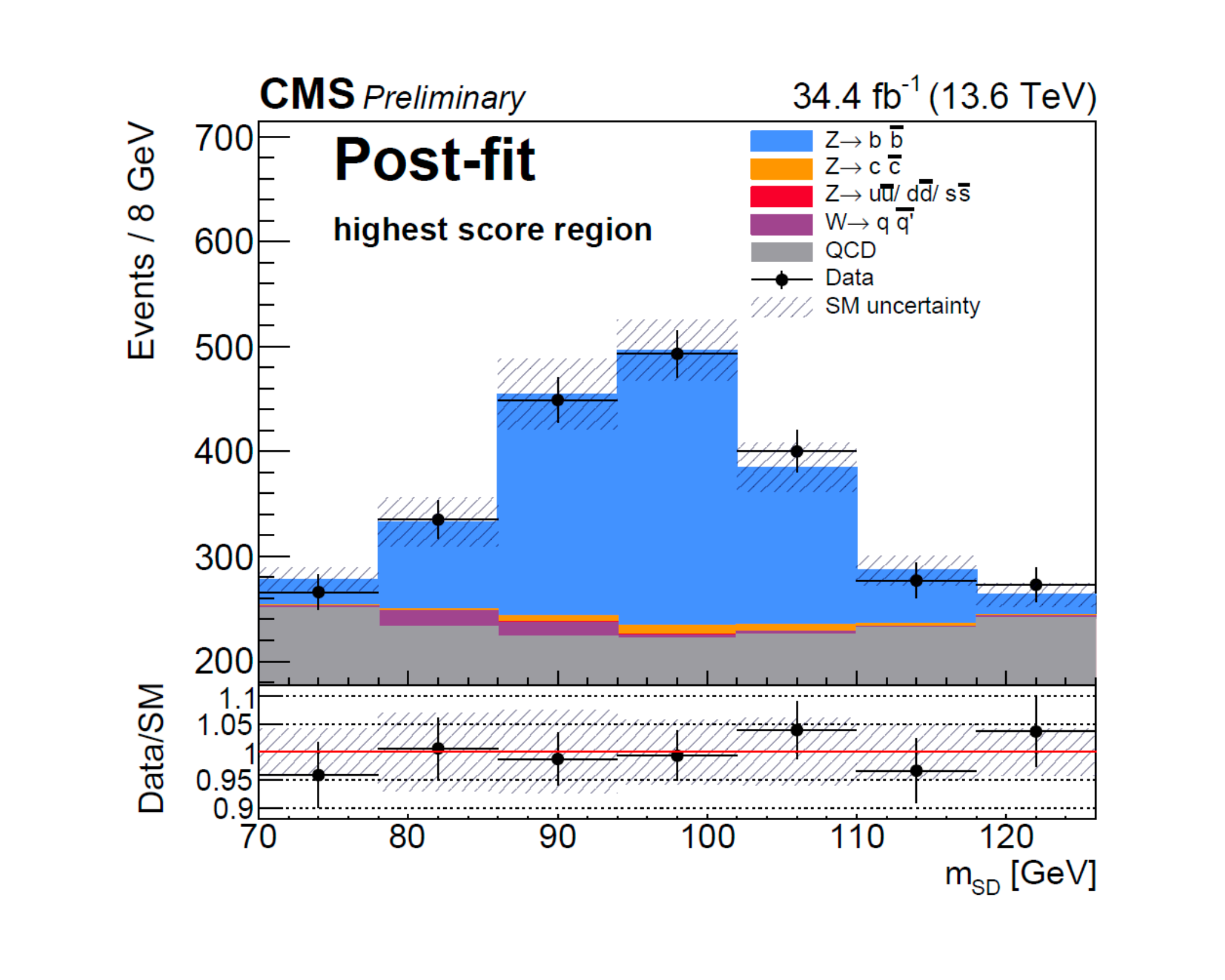}
    \caption{Left: The ROC curve of the ParticleNetMD discriminant obtained using simulation (blue) with the three working points pointed out in simulation (filled circles) and in data (hollow circles), under the 2018 data-taking conditions with $\mathrm{p_T}$ > 450 GeV. Right: Data to simulation comparison in the ParticleNetMD highest score region. The upper panel shows the mass distributions of data (black markers) and prediction (stack of QCD, $W\rightarrow qq'$ and signal $Z\rightarrow q\bar{q}$ , with the latter split on the basis of its decay products). The lower panel shows the ratio between data and prediction. The Z-peak is visible in the data distribution.}
    \label{fig:08}
\end{figure}

The left plot of figure 8 shows the mistag rate versus the signal efficiency for $X\rightarrow b\bar{b}$  tagging in Run-2 at various working points (Tight, Medium and Loose) along with the results of a dedicated calibration. In Run-3, a validation is performed in $Z\rightarrow b\bar{b}$ enriched samples with 2022 dataset, where the main backgrounds include $W\rightarrow qq'$ and QCD~\cite{CMS-DP-2024-055}. The main background consists of QCD multijet and is estimated using data driven techniques. The events are categorized in five regions of ParticleNetMD score and a simultaneous likelihood fit is performed using the soft-drop mass ($\mathrm{m}_{\mathrm{SD}}$)~\cite{Larkoski:2014wba} distributions of the four highest ParticleNetMD score regions. The right plot of figure~\ref{fig:08} illustrates the post-fit mass distribution in the highest score region of the ParticleNetMD tagger, showing a pronounced Z boson mass peak which confirms excellent agreement with SM predictions. This consistency is critical for searches and measurements involving boosted jets.


\section{Frameworks}
In CMS experiments, efficient frameworks for jet tagging and other particle physics analyses are essential for handling large datasets and ensuring efficient processing. The frameworks used for training and commissioning workflows, enable fast and scalable solutions, particularly for tasks like flavor tagging in jets. 
\subsection{b-hive (Training Framework):}
The b-hive~\cite{CMS-DP-2024-020} training framework focuses on efficiently training machine learning models for tagging tasks. It uses ROOT~\cite{rene_brun_2020_3895860} filelists as input and automates the dataset construction, training, and inference tasks. The training framework is defined using a workflow pipeline that supports parameterization, making it adaptable for different use cases. Its modular design allows users to easily introduce custom models and tweak training parameters, such as dataset version, number of epochs, and threads for parallel processing.

\begin{figure}
    \centering
    \includegraphics[width=0.55\linewidth]{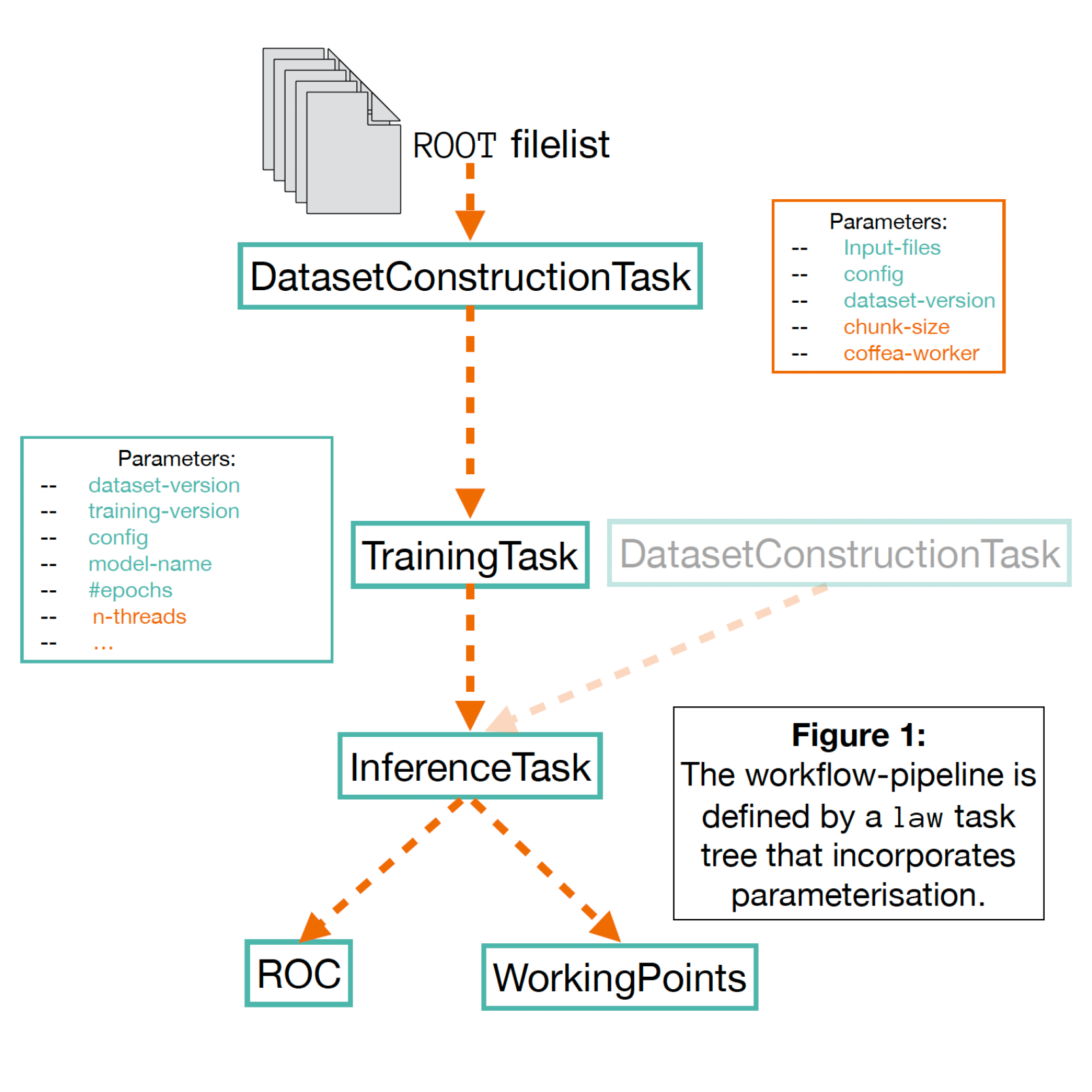}
    \caption{The flowchart of different stages in the b-hive framework.}
    \label{fig:10}
\end{figure}

This framework is highly customizable and flexible, enabling users to modify it to incorporate new models or methods, ensuring it can adapt to the evolving demands of jet tagging and related tasks within CMS. The efficiency and flexibility of this framework make it an invaluable tool for both model training and inference.
Figure~\ref{fig:10} shows a flowchart showing the workflow of the b-hive framework. 

\subsection{BTVNanoCommissioning~\cite{CMS-DP-2024-024}}
This workflow utilizes the NanoAOD format, which is a highly compressed data format containing essential information for physics analyses. After processing the RAW detector outputs into NanoAOD, the BTVNanoCommissioning framework is utilised for a faster, more efficient manipulation of data. It uses pythonic array-based manipulations rather than loops, streamlining operations by utilizing libraries such as Columnar Object Framework For Effective Analysis (coffea)~\cite{gray_2024_10977418}, awkward~\cite{jim_pivarski_2020_3952674},  and ROOT.  The automation is enhanced by GitLab Continuous Integration (CI/CD), enabling consistent performance monitoring over regular intervals with minimal human intervention.

\begin{figure}
    \centering
    \includegraphics[width=1\linewidth]{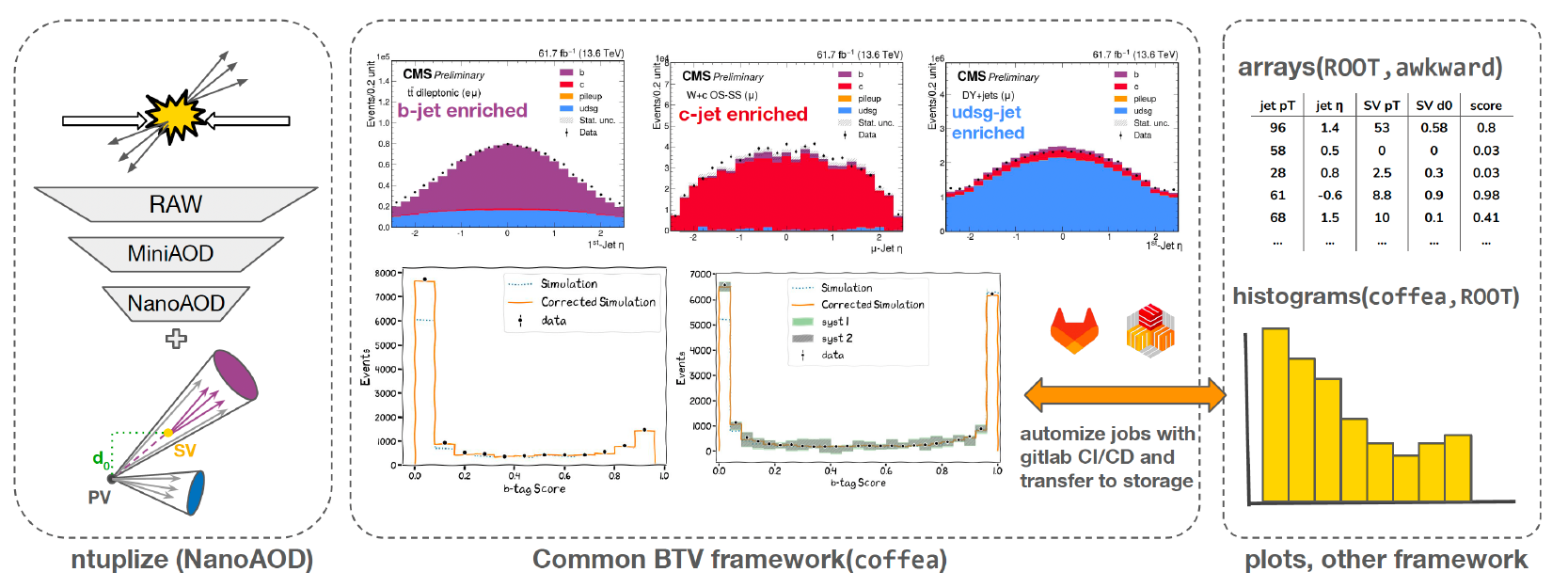}
    \caption{The analysis workflow for BTV commissioning tasks. The first stage is centrally produced NANOAOD samples, the second stage is divided into different workflows enrich in specific flavored jets, and the last part is the histogrammer and plotter. The last two stages are connected to gitlab CI/CD for automatically submitting jobs and producing the comparison plots.}
    \label{fig:11}
\end{figure}

Additionally, the BTVNanoCommissioning framework brings together selections from different phase spaces and applies corrections and systematic variations. It also has options of converting outputs into arrays, histograms, and figures and automatically transfers them to the storage area (EOS) via GitLab CI pipeline.. The framework helps ensure accurate monitoring of the jet tagging process and simulations for heavy-flavor jets. 
Figure~\ref{fig:11} shows the diagram of the full chain of processes that are executed for the production of commissioning results. 

\section{Summary}
The latest advancements in heavy-flavor jet tagging in the CMS experiment significantly improves the performance during Run 3 data taking in CMS.
Recent tagging algorithms in CMS utilises sophisticated machine learning techniques, including deep neural networks, graph-based architectures, and transformer models like the UnifiedParticleTransformer (UParT) algorithm. These models enhance the ability to distinguish between heavy-flavor jets and light-flavor or gluon jets. The proceeding reviews the evolution of these algorithms, emphasizing the improved tagging performance and reduced mistagging rates, along with the introduction of robust calibration techniques to mitigate discrepancies between data and simulation. Additionally, UParT also introduces novel techniques for s-quark jet tagging and hadronic-tau jet classification. 

The proceeding also discusses the performance of tagging algorithms for boosted jets and their importance in specific physics processes, such as $H \rightarrow b\bar{b}$ and $H \rightarrow c\bar{c}$.
The calibration of tagging performance using scale factors, derived from data and simulations, ensures accurate physics measurements. 

Finally, the paper outlines the two main frameworks used in CMS for model training and commissioning: b-hive, which automates machine learning model training, and BTVNanoCommissioning, which employs a compressed data format and automation for efficient analysis of data and simulations discrepancies.
\bibliographystyle{unsrtnat}
\bibliography{refs}


\end{document}